%% file: main.tex
\documentclass{sig-alternate}
\usepackage{mathptmx} 
\usepackage{pifont}
\usepackage{fancyhdr}
\usepackage{enumitem}
\usepackage[normalem]{ulem}
\usepackage[hyphens]{url}
\usepackage{tikz}
\usepackage[ruled,linesnumbered]{algorithm2e}
\usepackage[sort,nocompress]{cite}
\usepackage[final]{microtype}
\usepackage{flushend}

\newcommand*\circled[1]{\tikz[baseline=(char.base)]{
            \node[shape=circle,draw,inner sep=0.3pt] (char) {#1};}}
\usepackage[T1]{fontenc}
\usepackage[utf8]{inputenc}
\usepackage{authblk}
\usepackage{xspace}
\newcommand{\ours}{{Phoenix}\xspace}
\newcommand{\oursp}{{Phoenix$^+$}\xspace}

\title{Phoenix: Towards Persistently Secure, Recoverable, and NVM Friendly Tree of Counters} 
\author[1]{Mazen Alwadi}
\author[2]{Aziz Mohaisen}
\author[2]{Amro Awad}
\affil[1,2]{University of Central Florida}
\affil[1]{\textit {\{mazen.alwadi\}@knights.ucf.edu}}
\affil[2]{\textit {\{mohaisen, amro.awad\}@ucf.edu}}

\begin{document}
\maketitle
\pagestyle{plain}
\newcommand{\RNum}[1]{\uppercase\expandafter{\romannumeral #1\relax}}


\input{abstract.tex}

\input{intro.tex}

\input{background.tex}

\input{design.tex}

\input{evaluation.tex}

\input{related.tex}

\input{conclusion.tex}

\section*{Acknowledgement}
This research was developed with funding from the Defense Advanced Research Projects Agency (DARPA). The views, opinions and/or findings expressed are those of the author and should not be interpreted as representing the official views or policies of the Department of Defense or the U.S. Government. Approved for public release. Distribution is unlimited.



\bibliographystyle{IEEEtranS}
\bibliography{refs}

\end{document}

%% file: abstract.tex
\begin{abstract}

Emerging Non-Volatile Memories (NVMs) bring a unique challenge to the security community, namely persistent security. As NVM-based memories are expected to restore their data after recovery, the security metadata must be recovered as well. However, persisting all affected security metadata on each memory write would significantly degrade performance and exacerbate the write endurance problem. Moreover, recovery time can increase significantly (up to hours for practical memory sizes) when security metadata are not updated strictly.\par

Counter trees are used in state-of-the-art commercial secure processors, e.g., Intel's Safe Guard Extension (SGX). Counter trees have a unique challenge due to the inability to recover the whole tree from leaves. Thus, to ensure recoverability, all updates to the tree must be persisted, which can be tens of additional writes on each write. The state-of-art scheme, Anubis, enables recoverability but incurs an additional write per cache eviction, i.e., reduces lifetime to approximately half. Additionally, Anubis degrades performance significantly in many cases. In this paper, we propose Phoenix, a practical novel scheme which relies on elegantly reproducing the cache content before a crash, however with minimal overheads.  Our evaluation results show that Phoenix reduces persisting security metadata overhead writes from 87\% extra writes (for Anubis) to less than write-back compared to an encrypted system without recovery, thus improving the NVM lifetime by 2x. Overall Phoenix performance is better than the baseline, unlike Anubis which adds 7.9\% (max of 35\%) performance overhead.

\end{abstract}

%% file: intro.tex
\section{Introduction}

Non-Volatile Memories (NVMs) are emerging as promising contenders to the DRAM, and promise to provide terabytes of persistent data capacity that can be accessed using regular load and store operations\cite{Lee:2010:PTF:1749401.1749491,Li:2013:EHE:2495252.2495492}. Secure NVM systems commonly aim to protect confidentiality, integrity, and availability of the memory. While data persistency is an attractive feature that enables persistent applications, e.g., filesystems and checkpointing, it also facilitates data remanence attacks\cite{awad2016silent,chhabra2011nvmm,young2015deuce,ye2018osiris}. To protect the NVM data at rest, encryption becomes a necessity. Encryption targets the confidentiality among the security requirements. However, encrypting the data introduces the overhead of encryption metadata, which needs to be persisted to ensure secure and functional recovery \cite{ye2018osiris,ANUBIS,triad,liu2018crash}. In state-of-the-art secure processor systems \cite{awad2016silent,young2015deuce,costan2016intel,saileshwar2018synergy,awad2017obfusmem}, the counter-mode encryption is used due to its security and performance advantages.\par

To protect the integrity of encryption counters associated with each memory block (64 bytes), a Tree of Counters (ToC) or a Merkele Tree (MT) is usually used to verify the integrity of the counters themselves. ToCs are a form of integrity trees with security advantages over the MT since each counter in ToC is used in generating two different MAC values. Moreover, the trees of counters feature parallel updates due to their structure, which allows pre-computing all updates of a tree in parallel on each memory write. These advantages of trees of counters led to their deployment in commercial secure processors, e.g., Intel's Safe Guard Extension (SGX) support.

Unfortunately, due to the nature of these trees, they cannot be recovered from leaves. Thus, and unlike regular MTs, it is insufficient to just rely on persisting the leaves and rebuilding the tree after a crash\cite{ANUBIS}. Meanwhile, strictly persisting tree updates for each memory write can incur a significant write overhead, hence reducing the memory lifetime and significantly degrading the performance. Finally, rebuilding the tree after a crash can take hours if we are unable to identify the subset of tree nodes that have been possibly lost. Recent work \cite{ANUBIS} has demonstrated that a practical NVM size (e.g., 8TB) would require 7.8 hours for recovery. On the other hand, high-availability systems have stringent requirements of 99.999\% (five nines rule), i.e., the system can sustain a total of 7.8 hours down time only once each 89 years.\par

The state-of-the-art scheme, Anubis\cite{ANUBIS}, addresses the recovery time problem by persistently tracking the addresses of security metadata currently in the cache and thus limiting recovery of metadata to these addresses only. By doing so, Anubis no longer needs to rebuild the whole tree but only a subset of nodes that have been potentially updated and lost due to a crash. To enable ToC's recovery, Anubis uses a lazy update scheme while persisting each cache update to the NVM. Thus, Anubis only needs to verify that the shadowed cache (in NVM) has not been tampered with after a crash and then load that shadow cache content into the cache. By doing so, Anubis is able to recover the ToC but at the cost of a write to the shadow region on each security metadata cache update. Although  the first solution capable of recovering ToC integrity trees, Anubis' overhead can limit its deployment.\par 

NVMs' limited write endurance is perhaps the most challenging part towards its wide adoption  \cite{awad2016silent,awad2017obfusmem,young2015deuce,ye2018osiris,liu2018crash,towards,persistently}. In fact, encryption significantly exacerbates the write endurance issue due to the encryption's diffusion property \cite{young2015deuce}. Meanwhile, the state-of-the-art solution, Anubis \cite{ANUBIS}, incurs 87\% write overhead when used with ToC integrity trees, and systems using Anubis are expected to have almost half the lifetime span of systems without Anubis but no recoverability. Moreover, NVM writes are power hungry and have latency much higher than the read latency \cite{lee2009architecting,young2015deuce,chhabra2011nvmm}. Obviously, doubling the write bandwidth, as incurred by Anubis, limits its deployment and hence leaves NVMs in unrecoverable state.\par

In this paper, we aim to bridge the gap between recoverability and high-performance for secure NVM systems. We mainly focus on ToC integrity trees, due to its commercial adoption (Intel processors) and security advantages, including resistance to tampering and replay attacks. To bridge the aforementioned gap, we propose Phoenix, a novel memory controller design that achieves both recoverability and high-performance of secure NVM systems. Phoenix is based on our observation that reconstructing the exact content of the security metadata cache after a crash does not require shadowing all updates to the cache, i.e., allowing imprecise content addresses shadow of the metadata cache could be sufficient. In fact, we can reconstruct the exact lost cache state after recovery by recalculating the potentially lost values and then verify the integrity of the reconstructed cache. By relying on value recovery of the tree leaves only a small subset of updates to the cache need to be persisted in memory. Meanwhile, we still can verify that the recovered cache content reflects exactly the same cache state before the crash. Our optimization, realized in \oursp, relaxes persisting encryption counters on eviction, to only persist encryption counters on the N-th write, reducing \ours{}'s overhead significantly. 

While prior schemes, e.g., Anubis \cite{ANUBIS}, lazily-update the metadata cache which has its updates strictly shadowed to NVM, Phoenix allows the use of lazy-update metadata cache scheme but without strictly shadowing each update to the NVM. Phoenix is mainly based on four observations. First, most of the updates to metadata cache in a lazy-update scheme are for leaves. Second, leaf nodes are the least likely to be evicted, as they are going to be used frequently for integrity verification. Third, leaf nodes can be recovered using the encryption counter recovery schemes. Fourth, intermediate ToC nodes can be persisted at their location, and shadowing their addresses is sufficient for recovery, while ensuring the integrity by calculating the root of a small MT covering the cache. However, shadowing leaves updates to memory might be unnecessary if we can have the following: \circled{1} a mechanism to verify the most-recent cache state including leaves but without necessarily shadowing them, and \circled{2} the ability to recover leave updates. To do so, Phoenix elegantly employs relaxed counter recovery schemes which were used in the context of generic recovery of counters, but to recover the content of cache in a lazy-update scheme.\par

To evaluate Phoenix, we use Gem5\cite{gem5}, a full-system cycle-level simulator. By running representative workloads from the SPEC CPU2006 benchmark suite, we observe a reduction of write overhead from 87\% for Anubis---Phoenix has even less writes than write-back scheme, i.e., improves lifetime by almost 1.9x. Moreover, Phoenix has an average performance that is even less than write-back scheme, whereas Anubis incurs 7.9\% (max of 35\%) performance overhead.\par

The rest of the paper is organized as follows. First, in Section \ref{sec:back}, we discuss the background and motivation of our work. In Section \ref{sec:design}, we discuss the design of Phoenix. In Section \ref{sec:eval}, we discuss our evaluation methodology followed by our evaluation. We review the related work in Section \ref{sec:relat}. Finally, we conclude our work in Section \ref{sec:concl}.

%% file: background.tex
\section{Background and Motivation}
\label{sec:back}

In this section, we review background and related concepts, and motivate for our work. In particular, we start by defining the threat model, followed by all relevant concepts.

\subsection{Threat Model}
Similar to the state-of-the-art approaches \cite{yan2006improving,ye2018osiris,liu2018crash,rogers2007using,awad2016silent, gueron2016memory}, our threat model considers the processor chip to be the secure boundary. The processor contains the root of the integrity tree and the encryption key, where everything outside the processor is considered  untrusted. We assume an attacker is capable of performing passive and active attacks including bus snooping and replaying memory packets, can scan the memory contents, and may tamper with memory contents. We also assume the attacker can perform  attacks while the system is either on or off. Access pattern leakage attacks, electromagnetic (EM) inference attacks, and differential power analysis attacks are, however, beyond the scope of this work.

\subsection{Counter Mode Encryption}

\begin{figure}[t]
\begin{center}
\includegraphics[scale=0.7]{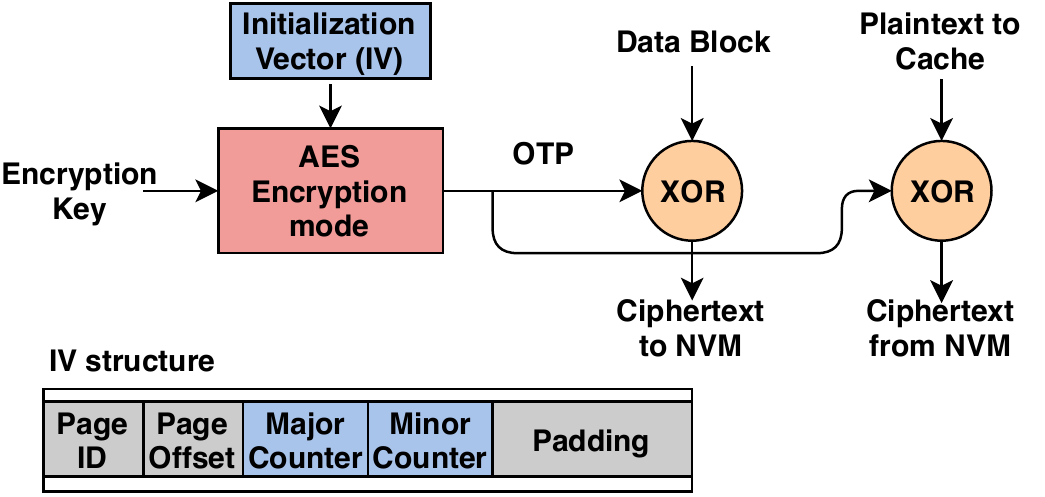}
\caption{Counter-Mode Encryption in state-of-the-art secure memories \cite{awad2016silent,ye2018osiris}.}
\label{fig:countermode}
\vspace{-2em}
\end{center}
\end{figure}

One of the major security vulnerabilities of NVM systems is the data remanence problem. Therefore, NVM is usually paired with encryption for data protection. The state-of-the-art secure processors (e.g., Intel Xeon Processor E-family) use counter-mode encryption, shown in Figure~\ref{fig:countermode}, since it provides strong defenses against a range of attacks (e.g., snooping, known plain-text, and dictionary-based). Moreover, the counter-mode has a smaller encryption/decryption overhead compared to other schemes due to overlapped latency of data fetching and one-time-pad generation\cite{awad2016silent,young2015deuce,awad2017obfusmem}. For each write to a data block, its associated counter will be incremented by 1. The updated counter is used to generate an initialization vector and then coupled with a processor key serve as inputs for the encryption engine to output a One-Time-Pad (OTP). After being XOR'ed with this OTP, the data block is encrypted and can be saved in memory. Similarly, a read request uses the same encryption pad to generate plain-text for processors but without updating any counter value. 

The size and organization of counters vary in different state-of-the-art schemes. The counters used in ToC are \textit{ \textbf{monolithic counters}}, where each of 56-bit long associated with a data block and one 64B counter cacheline can accommodate counters for eight 64B memory blocks. Encryption counter overflow can be costly, and causes long system stalls which is generally unacceptable\cite{rogers2007using}. Therefore, the monolithic counter should be large enough to prevent overflowing, which means more storage overhead.  For encryption/decryption, the monolithic counter will be padded with a block address to generate the initialization vector\cite{rogers2007using}.  To encrypt/decrypt the data, secure processors use AES counter-mode encryption. Figure \ref{fig:countermode} shows how counter code encryption works.

Several other state-of-the-art schemes use the \textit{\textbf{split counter}} scheme \cite{ye2018osiris,ANUBIS,triad,liu2018crash,saileshwar2018morphable,taassori2018vault,awad2016silent}, in which each data block is associated with one per-page major counter and one per-block minor counter. The major counter is shared by all the blocks within that page. Encryption/Decryption requires knowing both major and minor counter values to generate the OTP. Since each minor counter only accounts for seven bits, and the major counter for 64 bits, a small storage overhead occurs. However, when a minor counter overflows, the major counter is incremented by 1 and the whole page has to be encrypted using the new major counter\cite{ye2018osiris,ANUBIS,triad,liu2018crash,saileshwar2018morphable,taassori2018vault,awad2016silent}.

\subsection{Integrity Verification}
\begin{figure}[t]
\begin{center}
\includegraphics[width=1\columnwidth]{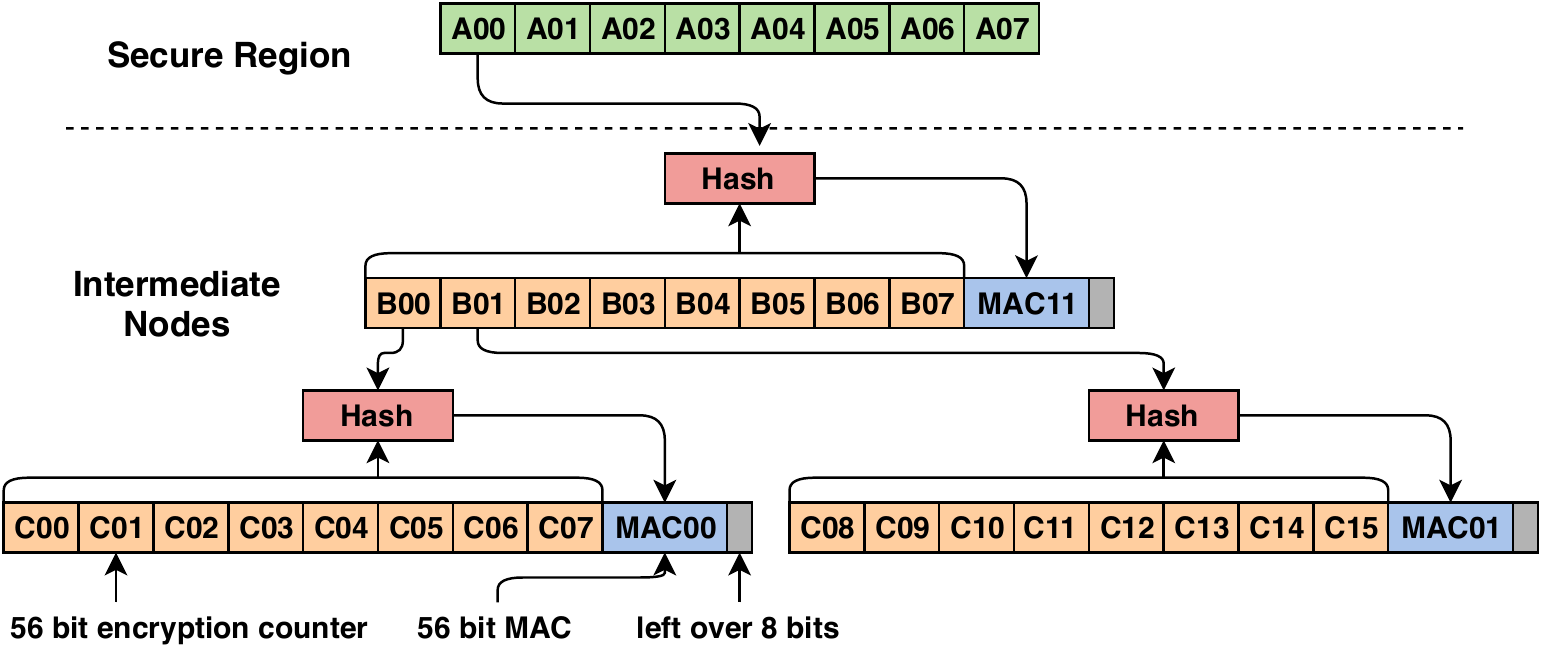}

\caption{SGX style Parallelizable Merkle Tree~\cite{gueron2016memory}}
\label{fig:SGXTREE}
\end{center}
\end{figure}

Since the trusted boundaries are limited to the processor chip, whenever a block is fetched from the memory, the memory integrity needs to be verified. In state-of-the-art research and secure processors design \cite{ye2018osiris,ANUBIS,triad,rogers2007using}, the Merkle Tree---one of approaches used for ensuring integrity, is widely studied and used for memory integrity verification.
Basically, Merkle Tree is an N-ary hash tree where its leaves correspond to encryption counters for data blocks \cite{rogers2007using} and every N leaves will have a hash value calculated based on the counter values. Similarly, all the intermediate nodes up to the root are constructed using the hash value based on its children. The root is always kept  secure; that is, it never leaves the chip. Moreover, any tamper with a counter leads to the failure of reconstructing the root. 
Depending on the tree structure, Merkle trees can be non-parallelizable (e.g., Bonsai Merkle Tree) or Parallelizable (e.g., SGX style counter tree) \cite{ANUBIS}. Since hashes in the Bonsai style trees are calculated over the bottom level hashes, the tree update must be done sequentially. ToC integrity trees, on the other hand, can perform a parallel update of the tree, as the MAC values are not calculated over the below level MAC value. Figure \ref{fig:SGXTREE} illustrates the organization of ToC integrity tree where each node is comprised of eight counters. The MAC values are calculated over these eight counters and one counter from the parent node as in Figure~\ref{fig:SGXTREE}. 

\subsubsection{Read and Verify}
To better understand the verification step in ToC integrity trees, Figure \ref{fig:SGXTREE} demonstrates a scenario of verifying counter {\tt C00}. Note that {\tt C00} falls within a block (64 bytes) that contains counters {\tt C00 - C07} in addition to a MAC value. However, verifying {\tt C00} also requires reading {\tt B00} in the upper level, and then calculating the MAC value over {\tt C00 - C07} and {\tt B00}, then compare it with {\tt MAC00}. However, it is important to note that this is assuming {\tt B00} is already verified and cached in the processor chip. However, if {\tt B00} is not present in the processor chip, it must be also verified the same way before we use it to verify {\tt C00}. 

Clearly, there is an inter-level dependency in the integrity tree, and missing an updated MAC due to a crash can cause  the whole recovery process to fail. 



\subsubsection{Write and Update}

To better understand how updates propagate through the ToC integrity tree, let's take the case of updating {\tt C00}. For now, let's assume that there is no expectation of integrity tree recovery after a crash. In its simplest form, updating (incrementing) {\tt C00} requires recalculating {\tt MAC00} after incrementing {\tt B00}. Similarly, {\tt MAC11} will be recalculated with the incremented {\tt B00} and {\tt A00} values. One important aspect to note here is that on each update, the MAC values on the affected nodes can be calculated in parallel using the incremented counter values. In contrast, and for regular Merkle Tree, calculating the upper levels requires the MAC value as an input, hence mandating the serialization of updates (bottom-up). Thus, ToC trees provide parallelism in updating the tree mainly because calculating the values of counters affected on each node can occur in parallel,  hence calculating the corresponding MAC values on each affected node.


\begin{figure*}
\begin{center}
\includegraphics[width=0.8\textwidth]{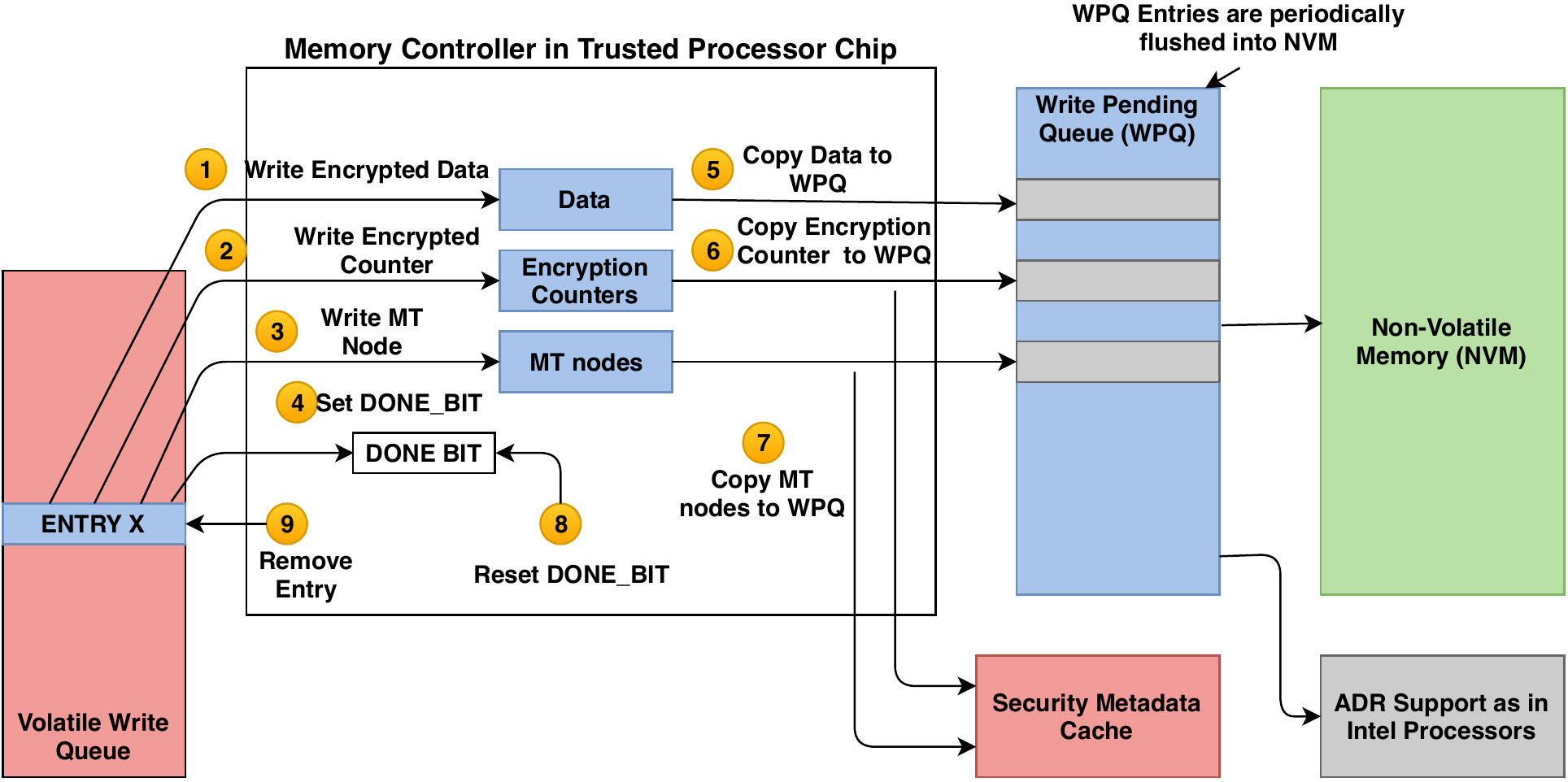}
\end{center}
\vspace{-1.7em}
\caption{Atomic Persistence of Integrity-Protected NVMs \cite{triad,ANUBIS}.}
\label{fig:TRIADNVM}
\end{figure*}

\subsection{ToC Advantages}
ToC provides security advantages over the Merkle tree, since in the ToC each counter/version is used in the calculation of two different MAC values, namely, the MAC value in the same node and the MAC value in the parent node. This, in turn, makes it harder to perform replay attacks. In ToC integrity-protected systems, the attacker needs to replay an old counter value that generates the same MAC value in the same node, and results in the same nonce in the parent node. However, in Merkle tree, the attacker only needs to replay a counter value that generates the same MAC value of the counter. For example, as shown in Figure~\ref{fig:SGXTREE} if an attacker wants to replay an old value of {\tt C03}, the value should generate the same MAC value as {\tt MAC00}, and should result in a value that generates the same MAC as {\tt MAC11}. Moreover, the new value of {\tt B00} should generate the same value in {\tt A00}, and the same process is repeated for all the levels of the ToC. In other words, to successfully perform an attack in a ToC integrity-protected system, the attacker needs to use a value that generates the same MAC values for all of the tree levels. On the other hand, for a Merkle tree integrity-protected system, the attacker only needs to use a value that generates the same MAC value in the parent node, then the parent MAC---which is basically the same, will be used to generate the upper levels. Moreover, ToC allows parallel calculation of upper levels updates, as the updates of upper levels does not depend on the MAC value of the child level. However, parallelism is not used when a lazy update scheme is used as we only write one cacheline and rely on eviction to propagate the updates to the upper levels.

\subsection{Metadata Cache Update Scheme and Recoverability in Persistent Memories}
Security metadata cache, caching the content of integrity tree and encryption counters, can be eagerly or lazily updated. In eager update schemes, each write needs to update all the related nodes up until the root in the cache. Thus, the root always reflects the most recent state of the tree and can be used to verify the memory integrity after recovery. In contrast, the lazy cache update scheme updates the leaf on each write and relies on propagating the updates upwardly only after the eviction of updated nodes. In the lazy update scheme, the root might still be stale while the metadata cache has the most recent values. Therefore, lazy update schemes are commonly paired with systems with no expectation of recovery\cite{triad,ANUBIS}.

In Merkle tree schemes where the tree can be regenerated using only the leaves (e.g., in General Merkle Tree), it requires a long time to rebuild the tree. Once the tree is reconstructed, the generated root will be compared against the one inside the processor chip which has been eagerly updated. In contrast, the lazy update scheme has no way to verify the integrity of the reconstructed tree since the root is out-of-date; the root does not reflect the most recent changes to memory before the crash. However, in the ToC integrity tree, it is impossible to regenerate the previous state of the tree from the leaf encryption counters since every intermediate node contains versions, the updated value of which could be lost during a crash. Due to such inter-dependency of levels, and volatile meta-data cache, it is very difficult to recover systems with ToC even if an eager update scheme is maintained.


While an eager update is suitable to rebuild the Merkle tree using only the encryption counters, it is not the case with ToC integrity tree. In ToC integrity tree, each node contains a MAC value calculated over the node counters and a nonce from the parent node. This inter-dependencies makes it very complex to retrieve the lost intermediate nodes during the recovery process. In lazily updated ToC integrity tree systems, the root is not enough to verify the integrity of the memory as it might be stale. Thus, to verify the integrity of the memory the integrity tree should be restored, while each node is used to verify the integrity of lower and upper levels.

\subsection{Counter Recovery Schemes}
Recovering encryption counters, the leaves of Merkle Tree, is generally considered the first step in recovering the tree. Prior work on general Merkle Tree \cite{triad, ye2018osiris} explored how to recover encryption counters after a crash. One solution, Osiris \cite{ye2018osiris}, relies on encrypting Error-Correcting Code (ECC) written with data. By limiting the number of updates to a counter before persisting it to memory, e.g., every 4th write, it can recover the counter used to encrypt the data by relying on the fact that a large number of errors will be detected by ECC when a wrong counter is used. By trying multiple counter values, Osiris can recover the counter used to encrypt the data. For more details on Osiris, the reader is referred to~\cite{ye2018osiris}.

While Osiris presents a novel approach that reduces the overhead of persisting counters significantly, there are many other competing approaches~\cite{liu2018crash}. For instance, as also discussed in~\cite{ye2018osiris}, part of the encryption counter used for encryption can be also written with the data and thus strict the persistence of the whole encryption counter can be relaxed. For the rest of this paper, we assume Osiris can be used, however, any other counter recovery scheme would work.

\begin{scriptsize}
\begin{table*}[t]
  \centering
  \begin{tabular}{|l|l|l|l|l|}
    \hline
    \textbf{Solution} & \textbf{ToC Recoverability} &\textbf{Low Recovery Time} &\textbf{Performance Overhead} &\textbf{Write Overhead}\\
    \hline
    \hline
      Osiris \cite{ye2018osiris} & X & Not Applicable & Not Applicable & Not Applicable  \\
     \hline
    Strict (TriadNVM \cite{triad}) & \checkmark & \checkmark & X (very high) & X (very high) \\
 
    \hline
    
    Anubis \cite{ANUBIS} & \checkmark & \checkmark & X (high) &  X (high)\\
    \hline
    \textbf{\ours} & \checkmark & \checkmark & \checkmark (low) & \checkmark (low) \\
    \hline
  \end{tabular}
  \caption{Schemes comparison}
  \label{table:formatting}
\end{table*}
\end{scriptsize}
\subsection{ToC Recovery Schemes}

The state-of-the-art scheme, Anubis \cite{ANUBIS}, is the first to enable recovering ToC trees without strictly persisting all tree updates to the memory. Anubis mainly builds upon the observation that it is sufficient to just recover the state of the cache before a crash, even if a lazy cache update scheme was used. In other words, if we can recover the content of the metadata cache after a crash, to the same content before the crash and verify it, then it is sufficient. To do the cache recovery part, Anubis persists security metadata cache updates in the memory in a region called the shadow region, i.e., writing cache updates additionally to memory. However, to fit the tag of the cache block and the block content into 64-byte cache lines, part of the counters (i.e., most-significant bits) are trimmed and their updates are persisted immediately. Meanwhile, to verify the integrity of the shadow cache after a crash, a small integrity tree protects the shadow region and uses eagerly updated Merkle tree. The root of the shadow region tree is updated on each cache update. However, the intermediate nodes do not need to be persisted; the shadow region tree is implemented using a general tree where the root can be generated from the leaves. Thus, after power restoration, Anubis reads the shadow region and then calculates the root of the shadow region and compares it with that inside the processor chip. Note that in Anubis there will be two roots inside the processor chip, one for the shadow region tree (regular tree) and the other one (ToC  tree) is for the rest of the memory. Moreover, Anubis keeps two integrity trees, one for the memory which is the ToC integrity tree, which is updated using the lazy update scheme, whereas the second one is the small general Merkle tree covering the shadow region, which is updated using eager update scheme.

Obviously, Anubis incurs almost double the write bandwidth: on each memory write, updating metadata cache needs to be persisted to memory.

\subsection{Atomic Update of Security Metadata}
\label{atomic}

While persisting the security metadata allows the system to recover after a crash, if the crash happens when the security metadata and data could not be both persisted will lead the NVM content to be inconsistent. To ensure the security metadata is consistent with the data, the update should be done atomically. Modern processors provide enough power to flush the content of the Write Pending Queue (WPQ) when a crash occurs, and the power to flush the WPQ content is provided by the Asynchronous DRAM Refresh (ADR) feature \cite{triad}. Therefore, all writes that have reached the WPQ are considered to be persisted. Additional bits, such as READY\_BIT or DONE\_BIT can be used to ensure the content of persistent registers are inserted atomically to the WPQ \cite{triad,liu2018crash}. Figure \ref{fig:TRIADNVM} shows how atomic updates are done, the encrypted data, encryption counter, and the updated MT nodes are moved to persistent registers, then the DONE\_BIT is set. After that the updates are moved atomically to the WPQ, then the data is written to the NVM, and finally the DONE\_BIT is reset and the entry is removed from the volatile WPQ.

\subsection{Motivation}

The main goal of \ours is to provide a practical solution for recovering ToC integrity trees. Table \ref{table:formatting} summarizes the features of the current solutions in contrast with \ours.
While Osiris \cite{ye2018osiris} provides an efficient counter (tree leaves) recovery scheme, it fails to recover the ToC integrity tree. As discussed earlier, ToC instances have inter-level dependence, which makes rebuilding a ToC from the recovered leaves impossible. Strict persistence schemes require persisting all tree updates in memory while using an eager update scheme, i.e., the root reflects the most-recent tree status. Thus, the strict persistence has a low recovery time, where there is no need to rebuild tree, although incurs a significant write overhead and result in performance degradation.\par

\begin{figure}[t]
\begin{center}
\includegraphics[width=\columnwidth]{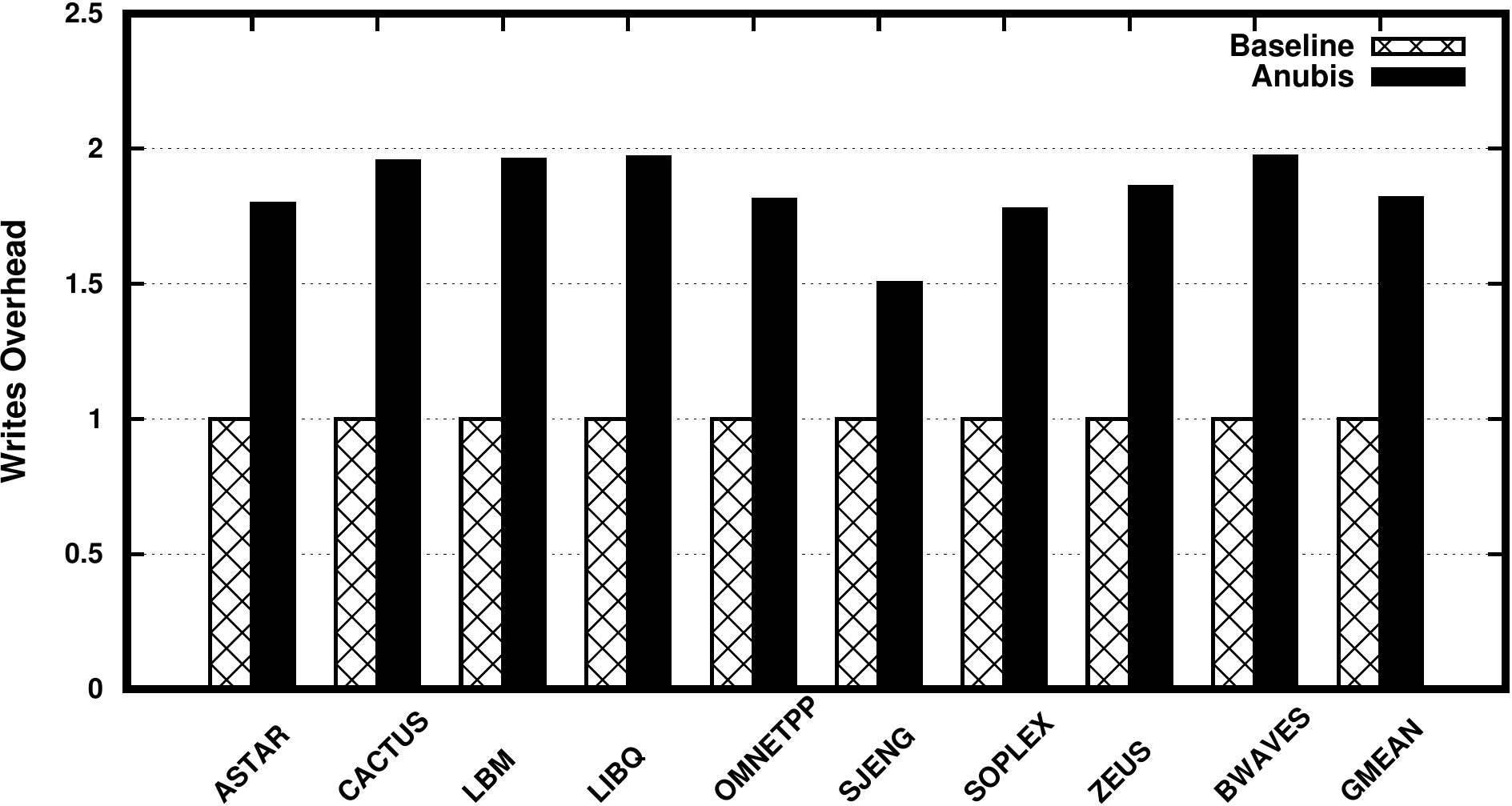}
\end{center}
\vspace{-1.7em}
\caption{Anubis extra writes.}
\label{fig:Motivation}
\end{figure}

For instance, for an 8TB memory system, strict persistence needs to persist additional 13 writes on each regular memory write, i.e., reducing NVM lifetime and increasing write bandwidth by 13x. Clearly, strict persistence is impractical. Anubis brings down the overhead of strict persistence significantly, although it is still high. In particular, Anubis incurs 2x the write bandwidth by persisting each update to cache in the shadow region. Thus, Anubis reduces the lifetime of NVM systems to almost half of its actual lifetime, although the lifetime of NVMs is already short, to begin with. Moreover, NVM writes are slow and power hungry, hence can significantly degrade the performance and increase the overall power consumption. Figure \ref{fig:Motivation} shows the overhead of Anubis scheme, which can limit its deployment, and motivates for this work. In particular, Figure \ref{fig:Motivation} shows the impact of Anubis on the number of writes. On average Anubis, incurs almost 2x the number of writes and average performance overhead of 7.9\% compared to baseline secure NVM without recovery support. The goal of \ours is to provide an NVM-friendly solution that does not incur significant NVM writes. Thus, \ours is proposed as a practical solution that realizes low-overhead secure and recoverable NVMs.

%% file: design.tex
\section{PHOENIX Design}
\label{sec:design}

Before delving into the details of \ours{}'s design, we first discuss the main goals and design principles that inspire \ours. The goal of \ours is to enable recovery of ToC integrity trees with an emphasis on low write overhead. We realize that any practical solution proposed for NVMs must have low write overhead. Thus, \ours mainly aims for ultra-low write overhead while still enabling recovery of ToC integrity trees. The first observation that \ours builds upon is that recovery of ToC integrity trees can be achieved by recovering the lost content of security metadata cache. While this observation has been also made in prior work, e.g., Anubis \cite{ANUBIS}, enabling such a recovery of cache content has been done in a way similar to write-through, by persisting the writes made to security metadata cache into a shadow region in the NVM, which has been proven to be very expensive when used with NVMs \cite{ye2018osiris}. However, \ours is based on the fact that we can actually recover the cache content without exact/accurate shadowing of all of its content. In this paper, we make a novel observation and contribution that we can securely recover the lost cache contents and verify them while still relaxing the shadowing operation. 

In particular, we observe that recovering ToC integrity trees by relying on restoring the cache content before a crash has two major requirements. First, there must be a mechanism to verify that we recovered the most recent cache content before a crash and its contents have not been tampered with. Second, the root of the Merkle Tree must reflect the updates of all memory including the cache contents just before a crash. By ensuring these two requirements are satisfied, the security metadata cache can be recovered and the rest of the memory verification is verified through a Merkle Tree on each memory access. In other words, simply bringing the metadata cache and Merkle Tree root (unaffected) to the state before a crash is sufficient to ensure crash consistency of security metadata.

Prior work, Anubis \cite{ANUBIS}, achieved such a cache recovery mechanism by relying on a reserved region in the NVM called the shadow cache, in which any updates of a lazy-update ToC metadata cache is copied, thus resulting in doubling the writes. To ensure the integrity of the shadow cache, Anubis applies a small Merkle Tree over the shadow cache while keeping its root in the processor and following an eager update scheme. After a crash, the cache content can be restored from the shadow region and its integrity can be verified using the small Merkle Tree (the eagerly updated one), which also has its root kept in an NVM (or NVM-backed) register inside the processor chip. On the other hand, \ours is mainly based on the fact that most updates to metadata cache in the lazy-update scheme are for leaves. However, shadowing leaves updates to memory might be unnecessary if we can have the following: \circled{1} a mechanism to verify the most-recent cache state including leaves but without necessarily shadowing them, and \circled{2} the ability to recover leave updates.   

\ours employs state-of-the-art counter recovery schemes, e.g., Osiris and phase-based recovery\cite{ye2018osiris}, to relax updates to the shadow region in the cache while simultaneously allowing to recover the exact content of the cache right before a crash. Specifically, \ours selectively decides which security metadata should be shadowed strictly and which ones can be relaxed. Even though it relaxes the shadow region update, \ours enables the reconstruction of the cache content (including relaxed leaves) and allows the verification of recovering the exact content before a crash. Since most updates to the security metadata cache are caused by leaves updates, \ours is expected to significantly reduce the number of writes while allowing fast recovery of ToC trees. The main downside of \ours is that it requires additional work before reconstructing the lost cache content and verifying it.      



The rest of this section discusses the design details of \ours. Note that the main distinction between \ours and other schemes lies in \ours{}'s ability to recover the exact cache content before a crash, although without a strict shadowing of cache content to NVM. \ours is the first scheme to enable lazy-update cache recovery without the need to persist each security metadata cache update.  


\subsection{Selective Persistence}
Upon a crash, the cache loses its contents. Losing the cached security metadata results in integrity verification failure, thus the cached security metadata needs to be persisted.

\begin{figure}[t]
\begin{center}
\includegraphics[width=1\columnwidth]{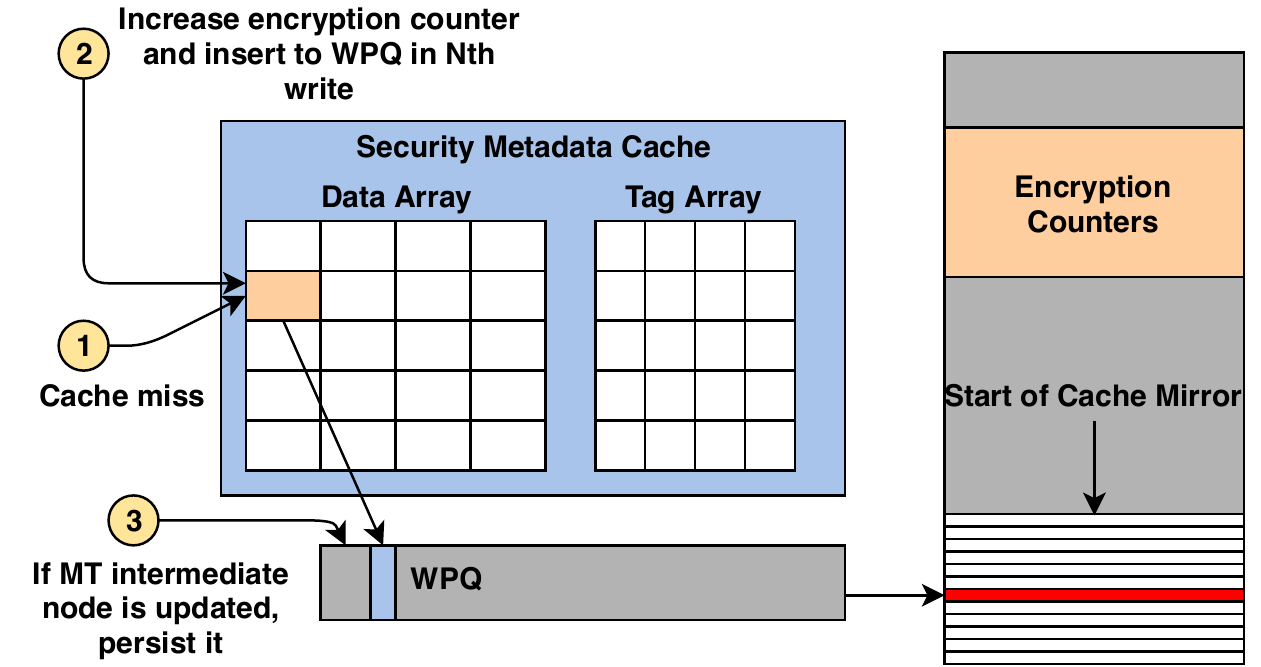}
\caption{Updates Tracking}
\label{fig:SHADOW}
\end{center}
\end{figure}

Strictly persisting the security metadata incurs tens of additional writes, and to avoid those unnecessary extra writes we opt for persisting only the unrecoverable nodes of the metadata. Security metadata contains the encryption counters and the Merkle tree nodes, while ToC integrity tree nodes are composed of 8 counters and a MAC calculated over the 8 counters and the parent of the node. Since the encryption counters can be retrieved without strictly persisting them using Osiris \cite{ye2018osiris}, we are going to follow a similar scheme by persisting the encryption counters with every N-th write, or on eviction. On the other hand, the intermediate ToC nodes are not recoverable, therefore we suggest persisting these cached nodes to successfully recover from the crash. To achieve that, we allocate a small region in the NVM which is the same size of the security metadata cache (about 256 kB) which we refer to as the Cache Mirror (CM). Whenever an intermediate ToC node is written in the cache, this update will be persisted and its address will be copied to the CM, while updating the encryption counters with every N-th write. Figure~\ref{fig:SHADOW} shows how selective persistence is done, whenever a write happens, if the write resulted in updating an intermediate node, the updated intermediate node is copied to the CM region. 
During the recovery process, the contents of the CM are used to recover the lost cache contents and refresh the ToC to ensure a secure recovery process. Since the security region is defined by the boundaries of the processor, the integrity of the CM should be guaranteed before it can be used during the recovery. Thus, we apply a small Merkle Tree (MT), four levels with an arity of eight, over the CM while keeping the root of this tree in the processor. During the recovery, the integrity of the CM region is verified by building the CM-MT and comparing the resulting root with the processor kept root.

\subsection{\ours Operation}
\ours read operation is merely a read and verify operation, and does not require any changes or special handling. In particular, the read operation in \ours does not modify the security metadata cache except for eviction, which is discussed in subsection \ref{subsec:eviction }. On the other hand, the write operation results in an encryption counter increment to ensure a new encryption pad for the modified block. The encryption counter increment will not affect the MAC value in the node nor increment the parent as we are using a lazy update scheme, but \ours will be triggered and the address of the modified counter will be copied to the CM. However, when an encryption counter block is evicted the parent node should be fetched and both nodes should be updated. Despite using a lazy update scheme, it is important to persist the encryption counter at every N-th (4th in \ours) write, or on eviction to enable encryption counter recovery. It is important to keep in mind that updates of ToC node and the data are to be done atomically using the Write Pending Queue (WPQ) and a ready bit as described earlier. While encryption counters are updated at every N-th write, ToC intermediate nodes need to be persisted each time they are modified, thus the addresses of the intermediate nodes are copied to the CM, and the intermediate nodes are persisted into the NVM.

\subsection{\oursp Operation}
While \ours persists intermediate nodes on each update, and persists encryption counters on N-th update or eviction. \oursp relaxes persisting encryption counters on eviction, to only persist encryption counters on the N-th write. By doing this, \oursp reduces the number of writes and the performance overhead significantly. \oursp relies on recovering the encryption counters while working, by utilizing the encryption counters recovery scheme. Notice that, recovering the encryption counters on the run might add performance overhead if done in a sequential manner, but we assume N-ECC engines (4 in our design) to retrieve the latest value of the used encryption counter. Keep in mind, that evicting an encryption counter without updating its value, does not affect its parent, but still affect the encrypted data. Thus, the old encryption counter value integrity can be verified the parent value, and the latest value can be recovered.

\subsection{Eviction}
\label{subsec:eviction }
The lazy update scheme we use in \ours reduces the number of writes while relying on eviction to propagate the nodes update. 
\begin{figure}[t]
\begin{center}
\includegraphics[width=\columnwidth]{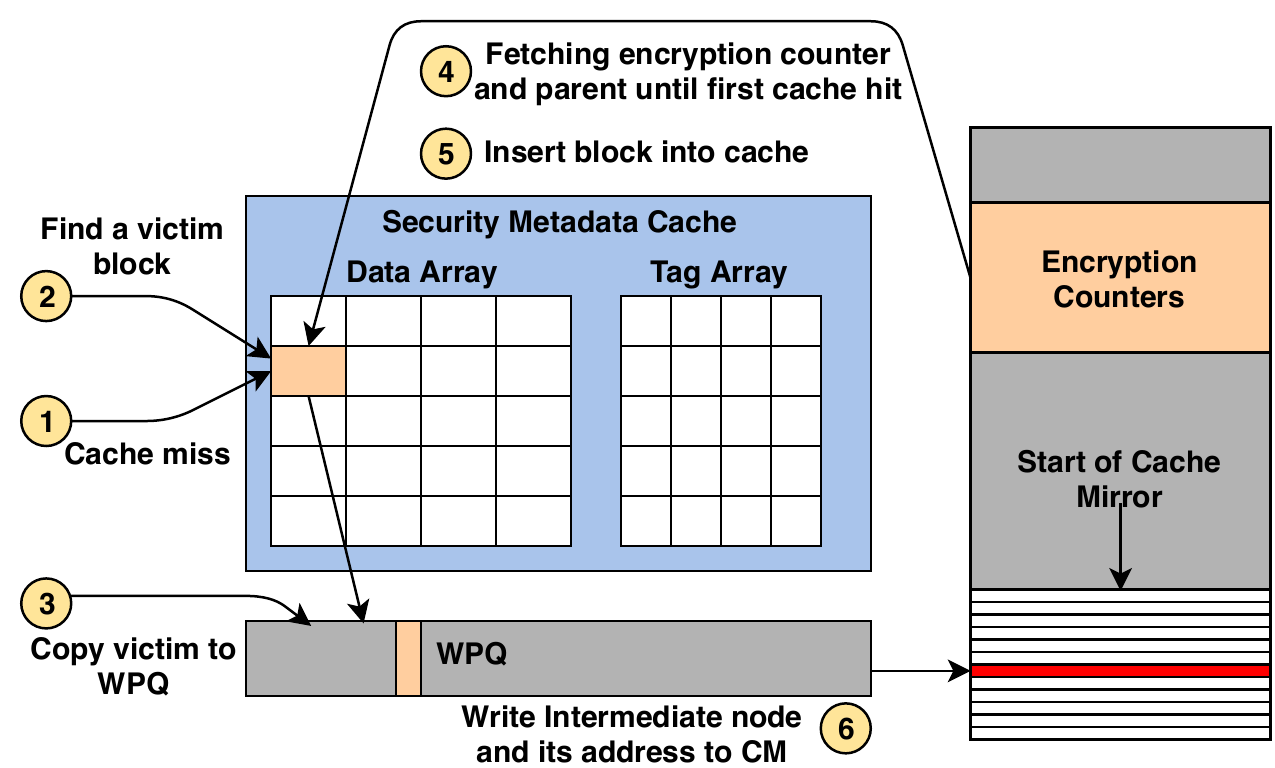}
\caption{Eviction}
\label{fig:Eviction}
\end{center}
\end{figure}
Figure \ref{fig:Eviction} shows the eviction process. In the case of a counter encryption cache miss, the memory controller selects a victim block to be evicted from the cache using the Least Recently Used (LRU) replacement policy. The victim block is then inserted to the WPQ in case it was an intermediate tree node. Note that we are not persisting the encryption counter on eviction, since we are relying on Osiris to retrieve the encryption counter's most recent value while running. Persisting the encryption counter on eviction will improve the performance slightly, as we can immediately use the fetched encryption counter after verifying its integrity, although will increase the number of writes to the NVM. Since our goal is to successfully recover the ToC while maintaining a low number of writes, we opt for recovering the encryption counter while running by only persisting the encryption counter at every N-th write. To ensure the data consistency we assume the evicted encryption counter, intermediate ToC node, data, and CM data are inserted atomically to the WPQ as described in section \ref{atomic}.

\subsection{Imprecise Cache Mirror}

\begin{figure}[t]
\begin{center}
\includegraphics[width=\columnwidth]{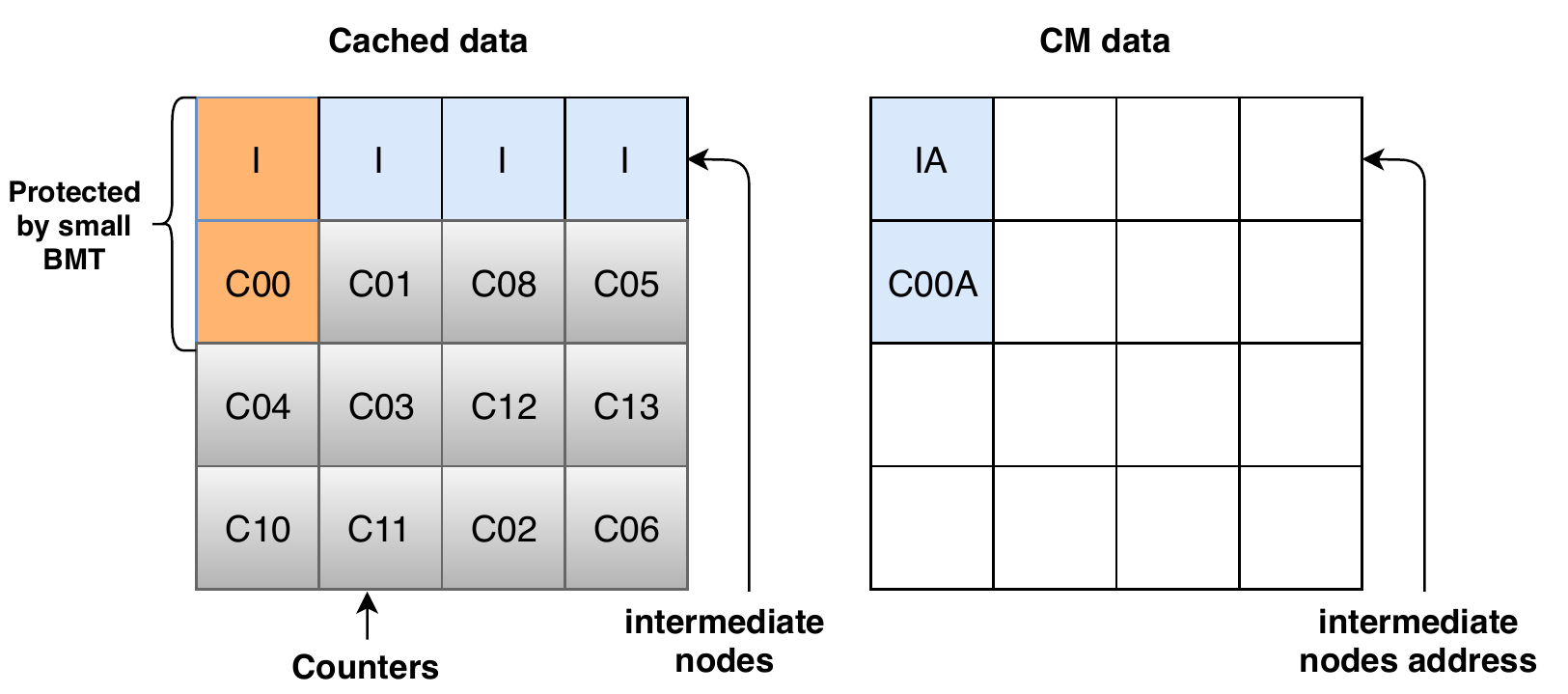}
\caption{Cache Mirror}
\label{fig:CacheMirror}
\end{center}
\end{figure}

The Cache Mirror (CM) region, shown in Figure~\ref{fig:CacheMirror}, is a small reserved region in the NVM. The CM only contains the addresses of the dirty intermediate nodes and the addresses of the  dirty counters, while the actual dirty intermediate nodes are persisted to their actual locations. The CM contents are used to securely recover the system after a crash. To ensure the integrity of  recovery, the dirty cached intermediate nodes and the dirty encryption counters are protected with a small general MT. This small MT that only covers the dirty intermediate nodes and dirty encryption counters is eagerly updated, and its root is always kept in the processor. Notice that by relaxing the CM contents to contain only the addresses of dirty intermediate nodes and  the addresses of dirty encryption counters, we were able to drop the number of writes significantly. Moreover, using a small MT to cover only the dirty intermediate nodes and dirty encryption counters, we were able to drop the performance overhead. We note that when a lazy update scheme is used, the root is no longer suitable as a single point of memory content integrity verification. As a matter of fact, the cache contents are the most updated nodes, and the nodes are used to verify the integrity of fetched nodes. When a crash happens, and the cached nodes are lost. However, we can recover the leaf nodes using encryption counters recovery, although the integrity of these nodes needs to be verified. The parents of these nodes can be either up-to-date in the NVM, or cached nodes lost during the crash. Thus, we make sure to persist the intermediate nodes, and use the small MT root to ensure the integrity of cached intermediate nodes.

\subsection{Integrity Verification}
The secure region is defined by the processors boundary. While on-chip memory is considered secure, that is not the case with NVM. Thus, whenever a data block is fetched from the NVM its integrity needs to be verified. To verify the integrity of any block, its parent needs to be fetched and used to calculate the MAC value of the verified block. However, once the parent is fetched, its integrity needs to be verified which will result in a recursive operation until the first parent cache hit. Once one parent is found in the cache, its integrity is considered to be verified, and is used to calculate the MAC of the child node. If the calculated MAC value matches the child node's stored MAC value, the child's integrity is considered verified. For the CM region, since its size is very limited (256 kB) it is more suitable to use an eagerly updated MT and store its root in the processor. Using an eagerly updating scheme means the root always reflects the most recent tree state. The CM MT is four levels using 8-ary tree, thus it is feasible to rebuild the tree during recovery and compare with the stored root to verify its integrity.

\subsection{Recovery}
The recovery process starts by loading the pre-crach cached intermediate nodes from the NVM, using the addresses saved in the CM region. Then, the integrity of the loaded intermediate nodes is verified using the small MT root. When the intermediate nodes are verified, any interrupted write operation is resumed, by checking the DONE\_BIT and completing the pending operations to successfully complete the atomic write. Notice that the small MT root is eagerly updated, and always kept in the processor. Moreover, the small MT root is calculated over the dirty cached intermediate nodes, thus its update is infrequent, since most of the updates are done to the leaf nodes. In turn, the overhead of eagerly updating the small MT root is negligible. Note that we are restoring the encryption counter during the normal operation, and the encryption counters are not persisted nor recovered during the recovery process, since they are recovered when fetched.
%

%

\begin{algorithm}
\SetAlgoLined
\KwResult{Recovered }
 Read CM content\;
 Load the intermediate nodes\;
 Load the encryption counters\;
 Recover the encryption counters\;
 Calculate small MT root and compare with processor root\;
  \eIf{small MT root = processor stored root}
  {
  \If{DONE\_BIT == 0}
   { 
    Move persistent registers content to WPQ\;
    Write WPQ contents\;
   }
   Proceed normal operation\;
   }
   {
   Failed to recover\;
  }
\caption{Recovery Algorithm}
\end{algorithm}

Once the CM integrity is verified, the DONE\_BIT is checked and any pending write operations that were in the persistent registers before the crash are moved to the WPQ and executed. After the pending write operations are executed, the CM contents are used to restore the cached intermediate ToC nodes. While the intermediate ToC nodes are ensured to be recovered to the most recent state using the CM, the encryption counters are not. To restore the encryption counter to the most recent state, we use the CM contents to retrieve the addresses of the cached encryption counters, then fetch the counters and use Osiris to retrieve the most recent counter value. After the encryption counters are updated to the most recent values, the cache is restored to its previous state before the crash, and its integrity can be verified using the small MT root. Notice that in case of the CM region is tampered with, and the calculated root of the CM region does not match the stored root in the processor, the recovery process fails and the integrity of the NVM is declared unverifiable.
\begin{figure*}[htbp!]
\begin{center}
\includegraphics[scale=0.7]{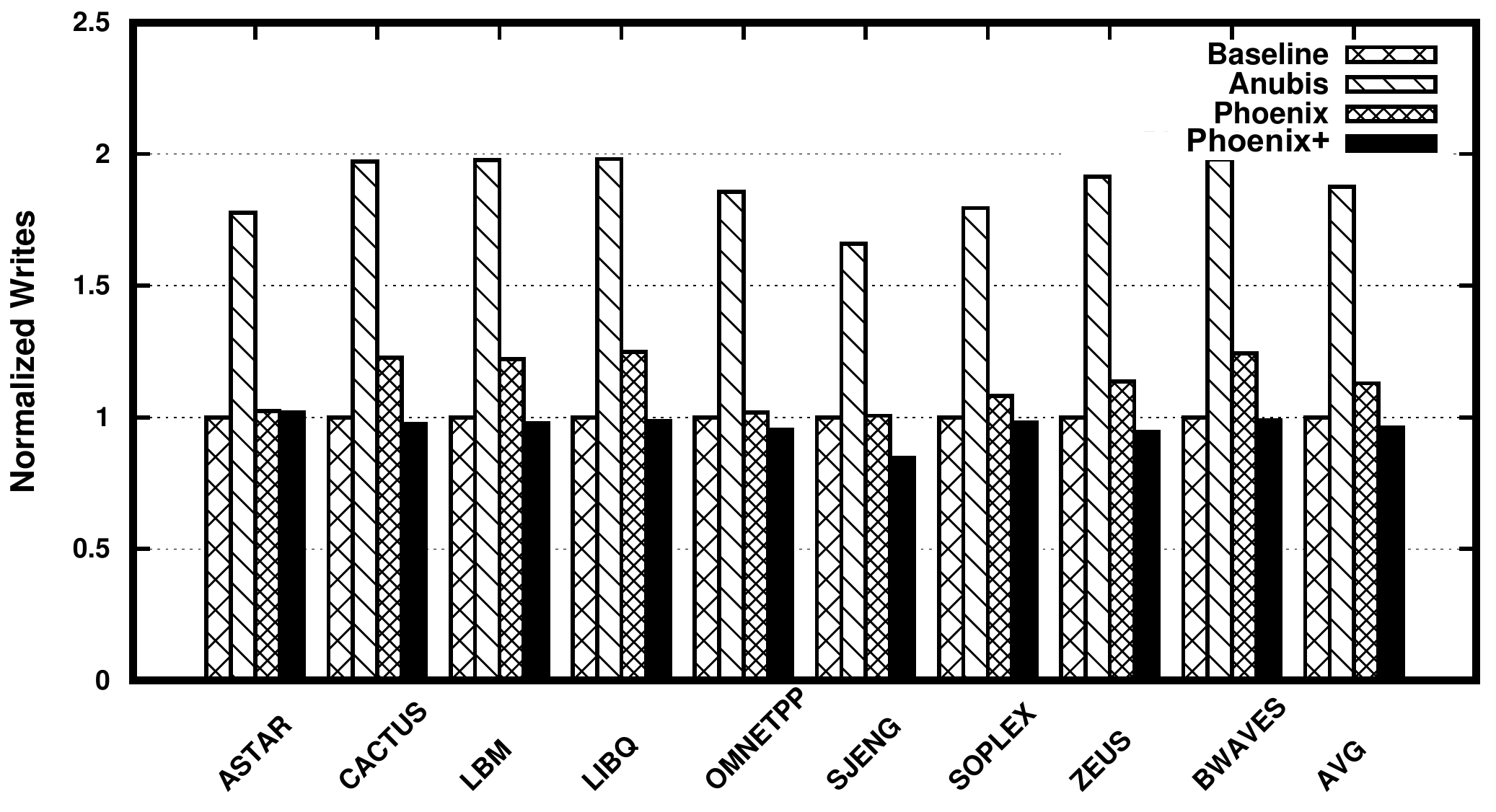}
\caption{Phoenix Extra Writes}
\label{fig:Writes}
\end{center}
\end{figure*}
\subsection{Security Discussion}
In traditional persistent secure systems, the security of the data is protected using the counter mode encryption, and the integrity of the encryption counters is protected with MT. The root of the MT is always kept in the processor, and memory content's integrity is verified by calculating the root and comparing it with the processor stored root. This scheme works well for eagerly updated MT, which is not the case in our scheme. \oursp scheme relies on a lazy update scheme, which means whenever a leaf counter is updated or evicted we do not update the parent of the counter, nor update the associated MAC with the leaf counter node, but rely on the N-th write to the same counter to propagate the update. In Figure~\ref{fig:SGXTREE}, if the counter {\tt C01} is updated twice and then got evicted, the parent node {\tt B00} will not be updated, and the MAC value {\tt MAC00} will both be stale. Notice that even the counter {\tt B00} will be stale in the NVM, so the next time counter {\tt C01} is fetched it still can be verified successfully using the stale {\tt MAC00} and the parent {\tt B00}, and then its most recent value can be recovered using Osiris \cite{ye2018osiris}. In the lazy update scheme, the root of the ToC can be stale, and the most updated state is preserved in the cached intermediate nodes of the ToC. In the recovery process, it is essential to guarantee the integrity of the CM region as it reflects the most recent state of the tree. Thus, the integrity of the CM is protected using a small BMT and the root is eagerly updated and kept in the secure region (processor).

%% file: evaluation.tex
\section{EVALUATION}
\label{sec:eval}
In this section, we evaluate our scheme based on the {\tt Write Back scheme} as the baseline, and compare it with state-of-the-art scheme, Anubis \cite{ANUBIS}. We evaluate the additional number of write incurred by each scheme, the performance of \ours and \oursp schemes, then we show the sensitivity to cache size and recovery time.
\begin{figure*}
\begin{center}
\includegraphics[scale=0.7]{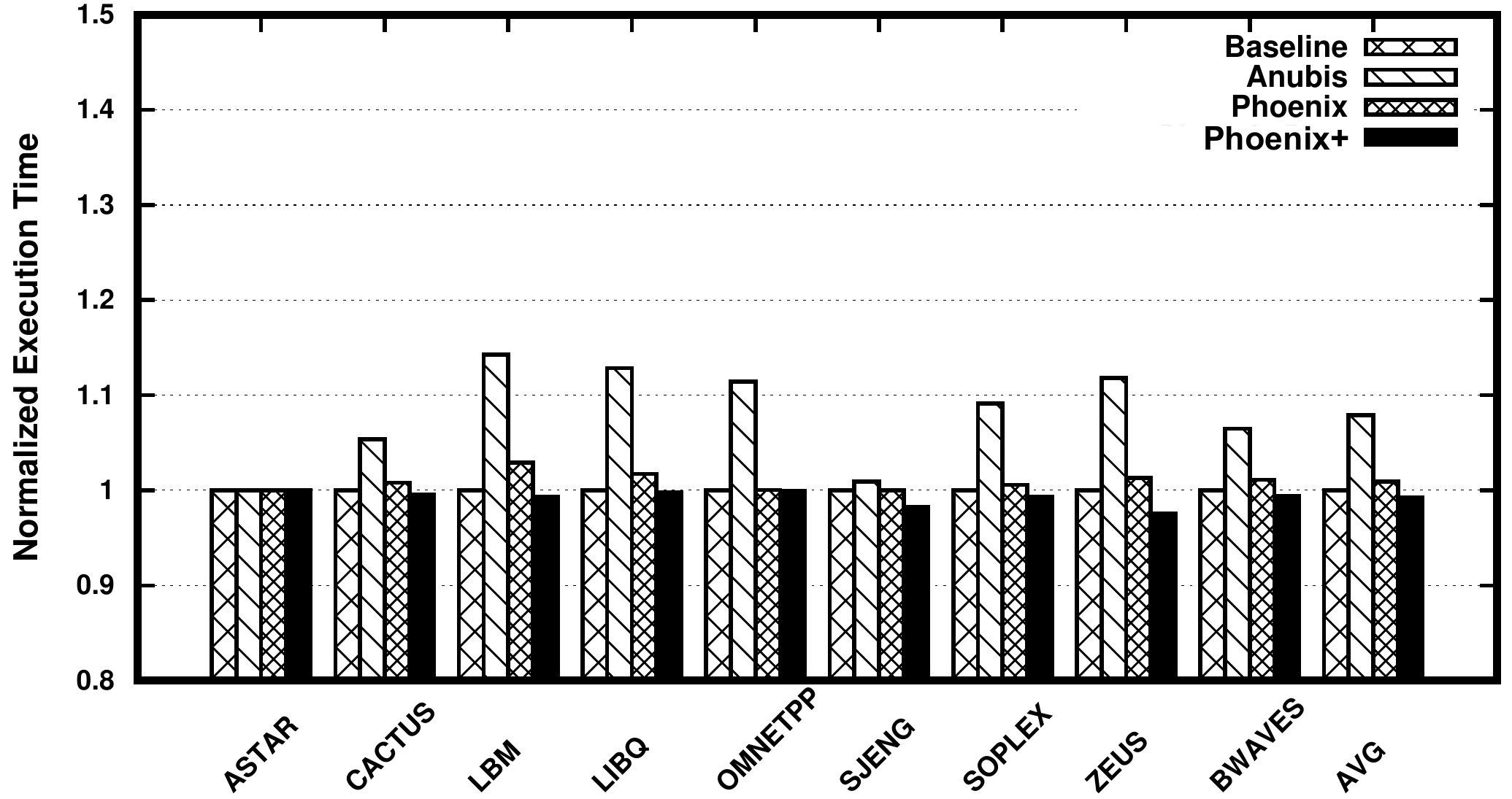}
\caption{\ours Performance}
\label{fig:Performance}
\end{center}
\end{figure*}

\subsection{Methodology}
\label{subsec:metho}
Evaluating our scheme was done using Gem5 simulator\cite{gem5}, a cycle-level simulator. Table~\ref{tab:configuration} shows the used configuration, we simulate a 4-core X86 processor with 16GB PCM-based Main Memory with parameters modeled as in \cite{lee2009architecting}. The evaluation was done by running 9 applications from the SPEC 2006 benchmark suit \cite{Henning2006}. The used benchmarks include memory intensive applications in both read and write intensive applications. For each application, we simulate 500M instructions after fast forwarding to a representative region.

\begin{table}[htp!]
 \vspace{-1em}
\centering
\caption{Configuration of the simulated system}
\label{tab:configuration}
\scriptsize
\begin{tabular}{|l|p{5cm}|}
\hline
\multicolumn{2}{|c|} {\bf Processor} \\ \hline                     
CPU          & 4 Cores, X86-64, Out-of-Order, 1.00GHz           \\ \hline
L1 Cache      & Private, 2 Cycles, 32KB,2-Way    \\ \hline
L2 Cache      & Private, 20 Cycles, 512KB, 8-Way \\ \hline
L3 Cache      & Shared, 32 Cycles, 8MB, 64-Way   \\ \hline
Cacheline Size      & 64Byte   \\ \hline
\multicolumn{2}{|c|}{\textbf{DDR-based PCM Main Memory}}             \\ \hline
Capacity      & 16GB                                        \\ \hline
PCM Latencies & Read 60ns, Write 150ns                      \\ \hline
\multicolumn{2}{|c|}{\textbf{Encryption Parameters}}                 \\ \hline
Security Metadata Cache & 256KB, 8-Way, 64B Block                     \\ \hline
CM in \ours & 256KB\\ \hline
CM in \oursp & 256KB   \\ \hline
Persistence Limit & 4   \\ \hline
\end{tabular}
\vspace{-1.6em}
\end{table}

 In our evaluation, we model the integrity-protection using ToC, encryption aspects, security metadata cache, hash calculation latency, and cache mirror region integrity protection.

\subsection{\ours Writes}
To evaluate our scheme, we compared the number of writes incurred for each of the following schemes. 1) \textbf{Write Back (Baseline)}: This is a simple ToC integrity tree scheme with write back. This system only writes on eviction and does not provide any recoverability. 2) \textbf{Anubis SGX}: The Anubis scheme for ToC integrity tree updates the ToC lazily and writes all the ToC updates to a shadow region. 3) \textbf{\ours}: This scheme updates the ToC lazily while persisting the updates for intermediate ToC nodes, and relaxes the updates for leaf nodes until eviction or the counter is written N times. 4) \textbf{\oursp}: This scheme reduces the number of writes in \ours by only persisting the leaf nodes on the Nth write, and relies on a counter recovery scheme (Osiris \cite{ye2018osiris}) to recover the counters on the run.


Figure~\ref{fig:Writes} shows the number of writes incurred by the above schemes. Considering the Write Back as the baseline scheme, we notice that Anubis incurs an average of 87\% extra writes, while \ours incurs an average of 12.9\% extra writes, and \oursp reduces the writes to less than the write-back by an average of 3.8\%. \oursp reduces the number of writes to less than Write Back scheme while achieving the recoverability of ToC. \oursp achieves this reduction by utilizing the lazy update scheme for the ToC, and by eliminating the eviction writes for the encryption counter nodes, while using Osiris counter recovery to recover the latest value of the encryption counter each time it is fetched.

\subsection{\ours Performance}
To evaluate \ours, we model and compare the aforementioned four schemes. Figure~\ref{fig:Performance} illustrates \ours's performance in comparison to other schemes. Considering the Write Back scheme as the baseline, Anubis provides the ability to recover the ToC with 7.9\% extra performance overhead. \ours  is not only capable of recovering the ToC, but also achieves a performance of 0.8\% better than the write back scheme. That is, \oursp' performance is less than the Write Back scheme, thus \oursp reduces the overhead by 8.7\% compared to Anubis. For instance, we notice, also from Figure~\ref{fig:Performance}, that \ours in both versions is performing better than the baseline for \textit{CACTUS} benchmark. Moreover, using memory intensive benchmarks shows that \oursp performs slightly better than the Write Back scheme, while this difference is expected to be more noticeable with less memory intensive applications. \ours reduces the overhead by relying on lazily updating the ToC while persisting each update to the intermediate ToC nodes. Notice that relying on the lazy update reduces the frequency of updating the intermediate nodes until the leaf node is evicted. On the other hand, \oursp takes one step further to relax persisting the leaf counters on eviction and relies on Osiris as a counter recovery scheme to recover the counters while running. 


\subsection{Sensitivity Study}

\label{sensitivity}
\subsubsection{Recovery Time}

\begin{figure}[ht]
\begin{center}
\includegraphics[width=\columnwidth]{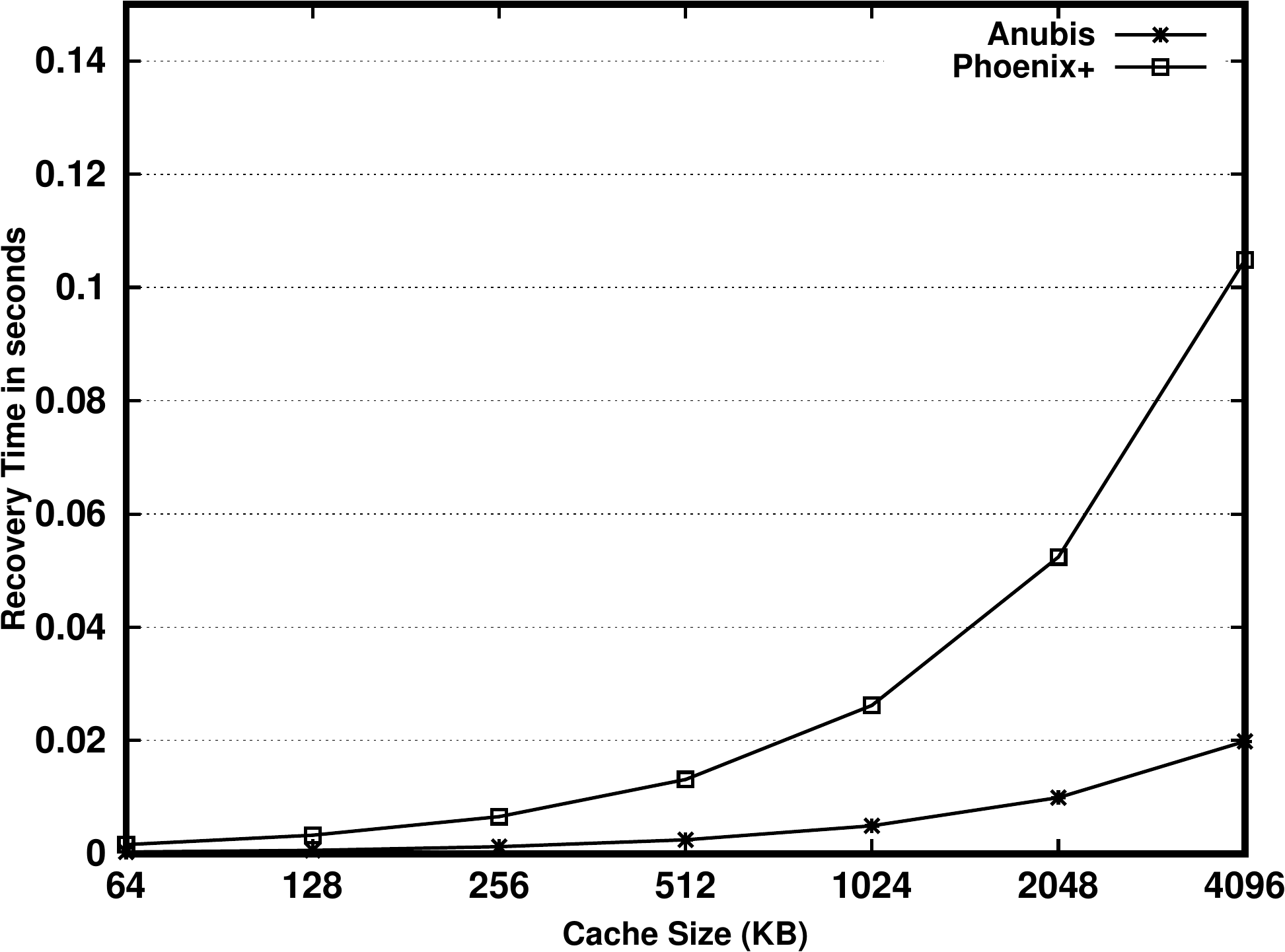}
\caption{Recovery Time}
\label{fig:Recovery}
\end{center}
\end{figure}
\vspace{-1.2em}
System recovery of ToC protected systems was not possible except for strict persisting scheme, until recently. Anubis \cite{ANUBIS}, \ours, and \oursp allow the recovery in less than a second, due to making the recovery time a function of the cache size instead of the memory size. While Anubis~\cite{ANUBIS} relies on a lazy strict persistent scheme which results in extra 87\% extra writes to achieve the recoverability of the ToC integrity protected systems in the same time as \ours requires to recover the same NVM. Figure \ref{fig:Recovery} shows the recovery time of Anuibs and \oursp regarding the cache size. Figure \ref{fig:Recovery} shows that both schemes achieve a recovery time of less than a second, even for extremely large cache size (4MB) \ours recovery time is $\approx$0.12 seconds for \oursp and $\approx$0.015 seconds for Anubis. Notice that \oursp recovery time is higher than Anubis scheme as it requires retrieving the leaf counters values during the recovery process. For this sensitivity test, the worst case scenario is considered to calculate the recovery time, by considering all the eight counters in each leaf node are stale. We notice that \oursp trades a very small amount of recovery time for reducing performance overhead and the number of writes of the system.

\begin{figure}[t]
\begin{center}
\includegraphics[width=\columnwidth]{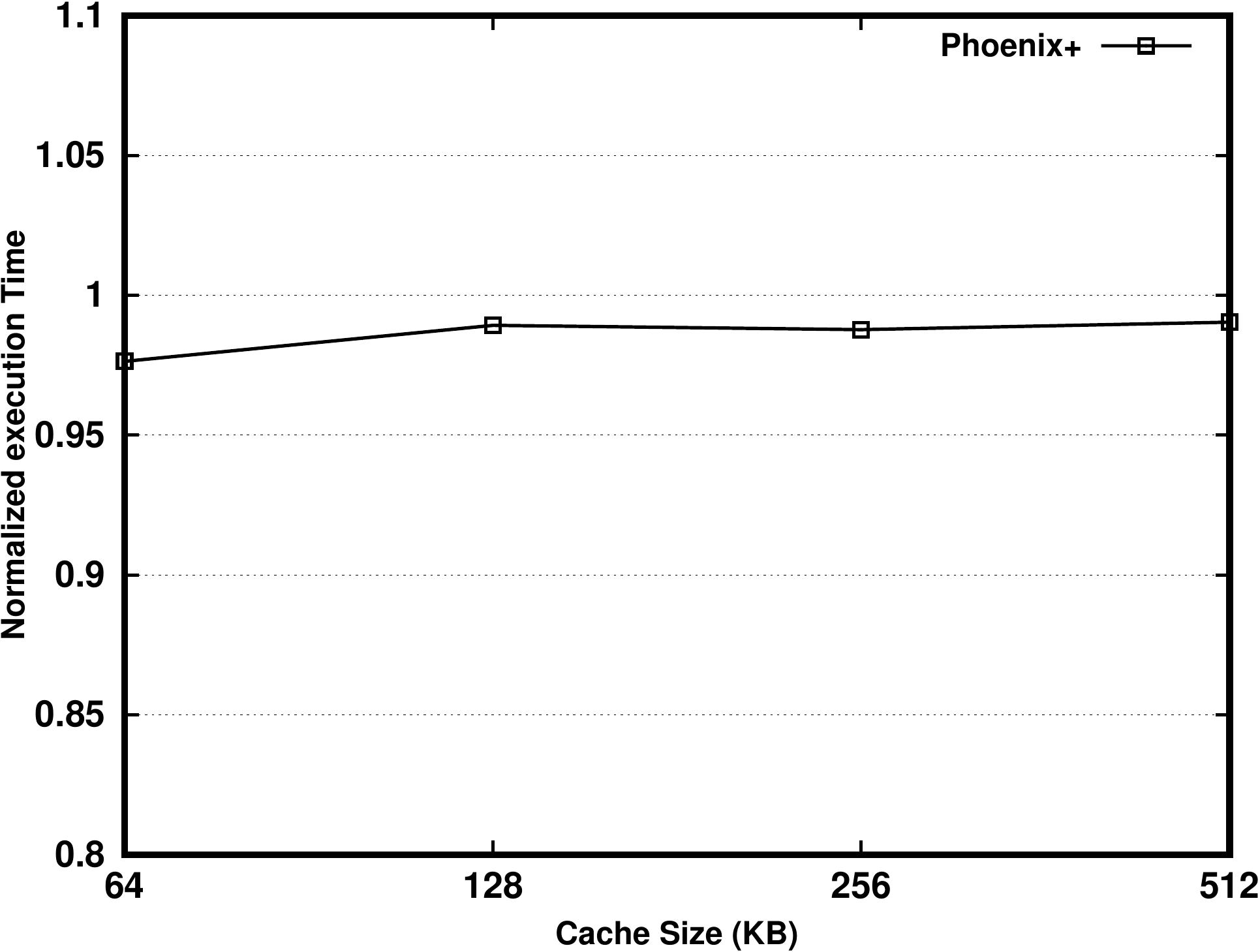}
\caption{Sensitivity to Cache Size}
\label{fig:sens}
\end{center}
\end{figure}

\subsubsection{Performance Sensitivity to Cache Size}
\oursp allows the recovery of ToC integrity protected NVM as a function of the cache size. To fully evaluate the scheme, we vary the cache size and measure the performance overhead of our scheme. As shown in Figure \ref{fig:sens}, the performance of \oursp almost stays the same. This can be explained by \ours operation; \oursp performs in a manner similar to the baseline (Write Back). However, \oursp' performance depends on the number of writes to cached data: the more writes to the cached data will result in more counter writes, which results in more \ours writes. 

\subsubsection{Counter Persistence Limit}

 The number of writes on which the encryption counter is persisted clearly affects the performance of \ours. Using a large number of writes before the encryption counter is persisted reduces the number of writes and the performance overhead. However, this comes at a cost: a large persistency limit would cause higher recovery time and higher performance overhead as the counter latest value needs to be recovered each time its fetched or during recovery. The performance overhead can be avoided by using multiple ECC engines to recover the counter value. In our design, we opt for using the 4th write to be the persistence limit, choosing to persist the counter at the 4th write provides a very low performance overhead and enables the recovery within less than a second.

%% file: related.tex
\section{RELATED WORK}
\label{sec:relat}
The most related work to \ours are Anubis\cite{ANUBIS}, Triad-NVM \cite{triad}, Osiris \cite{ye2018osiris}, and Crash Consistency \cite{liu2018crash}. Anubis \cite{ANUBIS} addresses the recovery time of NVM systems and uses a shadow region to track down all the changes to cache contents, where each writes to the cache results in a write to the shadow region. The shadow region facilitates recovering the cache contents in ultra-low time, but incurs 87\% extra writes for ToC. Triad-NVM \cite{triad}, on the other hand, discusses the trade-off between recovery time and performance, and reduces the recovery time by persisting N levels of the MT. On the downside, Triad-NVM does not work with ToC, and requires persisting multiple levels of the integrity tree.

To recover counters, Osiris \cite{ye2018osiris} relies on the ECC-bits as a sanity check for the used encryption counter. By applying a stop-loss mechanism, Osiris restores the encryption counters using a limited number of trials. Osiris works for retrieving the encryption counters while assuming building the integrity tree is possible, which is not the case with ToC integrity tree. Crash Consistency \cite{liu2018crash} for counters recovery proposes an API for programmers to selectively persist counters, and ensures atomicity through a write queue and Ready-Bit. In order to reduce the overhead, it proposes selective counter atomicity of the persistent applications. The scheme depends on the amount of applications persistent data and does not address the recoverability issue of ToC.

There are several state-of-the-art works done in NVM security and persistence\cite{coburn2012nv,zhao2013kiln,kolli2017language,taassori2018vault,saileshwar2018synergy,zuo2018secpm,saileshwar2018morphable,ren2015thynvm} without considering the crash-consistency and recovery that discusses to optimize the run-time overhead of implementing security to NVM. Most works employ counter-mode encryption for encrypting data and MT for ensuring integrity. However, to the best of our knowledge, none of the works consider the recovery and crash-consistency of integrity protected systems. As a matter of fact, any work that does encryption counters compression or increases the integrity tree arity boosts our scheme, by increasing the cacheability of the encryption counters and reducing the number of intermediate nodes. SecPM \cite{zuo2018secpm} proposes a write-through mechanism for the counter cache that tries to combine multiple updates of counters to a single write to memory, however, does not ensure recovery for ToC and incurs significant recovery time as in Osiris. While Anubis \cite{ANUBIS} discusses the reduction of recovery time of secure non-volatile memory and recovery mechanism for ToC, however; the scheme incurs almost 2x extra writes which reduces the NVM lifetime by half.

%% file: conclusion.tex
\section{CONCLUSION}
\label{sec:concl}
Phoenix is based on four observations, first, most updates of the lazily updated ToC are done to leaf nodes. Second, leaf nodes are the least likely to be evicted as they will be reused frequently for verification and update purposes. Third, leaf nodes can be recovered using any encryption counter recovery scheme, we used Osiris in our work, but any other scheme should work. Fourth, cached intermediate nodes can be persisted at their location instead of being copied to the shadow region, and the small MT only needs to cover the dirty cached intermediate nodes and the dirty encryption counters.
Phoenix achieves recoverability with ultra-low recovery time while keeping the number of writes to the minimum in ToC integrity protected NVMs. Our solution achieves a significant improvement in the number of writes as it reduces the number of writes by 90.8\% less than state-of-the-art scheme Anubis, and 3.8\% less than the write back scheme, with a recovery time of less than a second in ToC integrity protected systems. In addition, Phoenix recovery time and extra writes are a function of the cache size, as it works by recovering the lost cached ToC nodes. In summary, Phoenix recovers the ToC in less than a second, reduces the number of writes significantly, and improves the performance.